\documentclass{aa}
\pdfoutput=1
\usepackage{graphicx}
\usepackage{txfonts}
\usepackage{textcomp}
\usepackage{epstopdf}
\usepackage{subfigure}
\usepackage{array}

\newcommand{\ds}{\ensuremath{^\circ}}
\newcommand{\ms}{$\textrm{m\: s}^{-1}\:$}
\newcommand{\mss}{$\textrm{m\: s}^{-1}$}

\begin{document}

   \title{Collisions of small ice particles under microgravity conditions}


   \author{C. R. Hill
          \inst{1},
          D. Hei{\ss}elmann \inst{2,3},
          J. Blum \inst{2},
          H. J. Fraser \inst{1}
          }

   \institute{The Open University, Department of Physical Sciences, Walton Hall, Milton Keynes, MK7 6AA, UK\\
              \email{catherine.hill@open.ac.uk}
         \and
             Technische Universit{\"a}t Braunschweig, Institut f{\"u}r Geophysik und extraterrestrische Physik, Mendelssohnstra{\ss}e 3, 38106 Braunschweig, Germany
         \and
            International Max-Planck Research School, Max-Planck Institute of Solar System Research, Justus-von-Liebig-Weg 3, 37077 G{\"o}ttingen, Germany\\
             }

   \date{Received 25 April 2014 , Accepted 8 October 2014}


  \abstract
   {Planetisimals are thought to be formed from the solid material of a protoplanetary disk by a process of dust aggregation. It is not known how growth proceeds to kilometre sizes, but it has been proposed that water ice beyond the snowline might affect this process.}
   {To better understand collisional processes in protoplanetary disks leading to planet formation, the individual low velocity collisions of small ice particles were investigated.}
   {The particles were collided under microgravity conditions on a parabolic flight campaign using a purpose-built, cryogenically cooled experimental setup. The setup was capable of colliding pairs of small ice particles (between 4.7 and 10.8 mm in diameter) together at relative collision velocities of between 0.27 and 0.51 \ms at temperatures between 131 and 160 K. Two types of ice particle were used: ice spheres and irregularly shaped ice fragments.}
   {Bouncing was observed in the majority of cases with a few cases of fragmentation. A full range of normalised impact parameters (\textit{b/R} = 0.0-1.0) was realised with this apparatus. Coefficients of restitution were evenly spread between 0.08 and 0.65 with an average value of 0.36, leading to a minimum of 58\% of translational energy being lost in the collision. The range of coefficients of restitution is attributed to the surface roughness of the particles used in the study. Analysis of particle rotation shows that up to 17\% of the energy of the particles before the collision was converted into rotational energy. Temperature did not affect the coefficients of restitution over the range studied.}
   {}


   \maketitle
%

\section{Introduction}
\label{Introduction}
Since the first confirmed discovery of an exoplanet around a solar-type star in 1995 \citep{Mayor95}, well over 1000 exoplanets have been discovered. The sheer number of planets that have been discovered so far means that the question of how they form is of great scientific importance. Planets are thought to be formed from the material of the protoplanetary disk of gas and dust (around 1\% by mass) that surrounds a young star. Many of the initial models of planetesimal formation invoked gravitational instability in the dust, but it became apparent that turbulence in the disk would prevent this from occurring.

The theory of planetesimal formation by dust aggregation was first proposed by \cite{weidenschilling1977,weidenschilling1980} and has since gained widespread acceptance. Collisions of silicate dust particles have been studied extensively, in the laboratory, under microgravity conditions and in simulations, and sticking has been experimentally demonstrated with small particles at low collision velocities due to van der Waals forces (see reviews by \cite{blumwurm2008} and \cite{guettleretal2010}, and the more recent work by \cite{weidlingetal2012} and \cite{kotheetal2013}). There appears to be a critical velocity above which dust particles will no longer stick. This is around 1~\ms for micron sized particles and decreases with increasing particle size \citep{guettleretal2010,kotheetal2013}. However, direct growth between similar-sized particles beyond centimetre sizes has not been demonstrated because van der Waals forces no longer dominate in this size region \citep{Zsom10}. In addition, it was shown by \cite{weidenschilling1977} that relative velocities of particles in protoplanetary disks increase with increasing particle size making sticking unlikely for larger particles, which tend to bounce rather than stick \citep{guettleretal2010,Zsom10}.

This so-called "bouncing barrier" is a major problem with this theory and has been the subject of considerable experimental research. Various mechanisms have been proposed to overcome this problem. Collisional grain charging is one such mechanism \citep{PoppeBlum00, Poppe00}; however, this is only possible for small particles, so it is unlikely that planetesimals can form by this mechanism alone \citep{Blum10}. Growth has also been demonstrated through mass transfer \citep{wurmetal2005,teiserwurm2009a,teiserwurm2009b,Kothe10,teiseretal2011,windmarketal2012a,windmarketal2012b,garaudetal2013,meisneretal2013}, but owing to the increasing relative velocities of larger particles mentioned earlier and the importance of erosion \citep{schraeplerblum2011,seizingeretal2013}, it is unclear whether this can account for growth beyond centimetre sizes.

Recent research suggests that the effects of ice beyond the snowline may be able to overcome the bouncing barrier. It was suggested by \cite{Chokshi93} that dust coagulation in dense clouds requires the grains to be coated in an icy mantle for efficient sticking, and it is possible that same hypothesis will hold in planet-forming environments. Observational data point towards a substantial population of icy (water) grains  around young stellar objects (the precursors of planet-forming systems) in both crystalline and amorphous ice phases \citep{Schegerer10}. Further indirect evidence of large icy reservoirs of water in protoplanetary environments has been established from Herschel detections of cold water vapour (likely produced by photodesorption and sputtering) in outer disk regions, coupled with warm H$_2$O gas in denser planet-forming regions \citep{Dishoek09, Podio13}. \cite{Oberg11} show that the resulting C/O ratio in an exoplanetary atmosphere may be intrinsically linked to where the planet aggregated, in relation to the corresponding protoplanetary snowline. The first direct imaging of the CO snowline was recently reported, using N$_2$H$^+$ emission as a probe \citep{Qi13}. Therefore all the evidence points towards the presence of icy particles beyond the snowline in protoplanetary disks, and as such, their collisional properties must be taken into account when considering planet formation.

Ice particles have been found to have a larger rolling friction force \citep{Gundlach11} and reduced elasticity \citep{Hertzsch02}, which will increase the threshold velocity for sticking. In addition, recent model simulations have shown that ice condensation could enable dust grains to grow to decimetre sizes around the snowline \citep{Ros13} and to icy planetesimals if the initial ice grains were submicrometre-sized \citep{kataokaetal2013}. If the particles are composed of amorphous solid water, electrostatic effects may come into play, because it has been found that amorphous solid water can form permanent dipoles that would increase the sticking probability \citep{Wang05}.
To investigate the collisional properties and sticking of ice further, we have conducted microgravity experiments on millimetre-sized ice particles.

Key parameters used by modellers in the study of particle collisions are the coefficient of restitution, $\varepsilon$, and the impact parameter, \textit{b}. The coefficient of restitution is the ratio of the relative velocities of the particles after and before the collision:  \begin{equation} \varepsilon = \frac{v_{a}}{v_{b}} \end{equation} where \textit{a} and \textit{b} denote after and before the collision, respectively. It is related to the translational kinetic energy lost in the collision. The impact parameter is the distance of closest approach of the two particles perpendicular to their relative velocity vector. Henceforth we use the normalised impact parameter \textit{b/R}, which is the impact parameter, \textit{b}, normalised to \textit{R} which is the distance between the centre of masses of the two particles at the point of collision, i.e. the sum of the two radii. Work by \cite{Hatzes88}, \cite{Bridges84}, and \cite{Supulver95} investigated ice collisions utilising a disk pendulum and an ice target and found that coefficients of restitution decrease with increasing impact velocity. Frost on the ice surface also caused a reduction in the coefficient of restitution \citep{Supulver97}, while glancing collisions showed very little loss of kinetic energy. Free impacts of ice spheres (diameters 0.28-7.2 cm) with ice targets \citep{Higa96, Higa98} yielded a constant coefficient of restitution below a critical velocity, above which the ice began to fracture and coefficients of restitution decreased with increasing impact velocity.

To access  low collision velocities in free space (i.e. collisions without the use of a pendulum or dropping arrangement), it is necessary to conduct experiments under microgravity conditions. \cite{Heisselmann10} performed experiments on collisions of ice spheres 1.5~cm in diameter on a parabolic flight and found that there were a spread of coefficients of restitution from 0.0 to 0.84 for a spread of normalised impact parameter from 0.0 to 0.6 and relative impact velocities between 0.06 and 0.22~\mss. No correlation was found between coefficient of restitution and impact velocity.

In this paper, we present results of microgravity collision experiments of millimetre sized ice particles. We aim to investigate the coefficients of restitution, impact parameters and impact velocity to see if these parameters have any dependence on each other. In addition, we will present a number of special cases and comment on their relevance to planet formation scenarios.


\section{Experimental details}
\label{Experimental details}
The experiment was conducted on the German Space Agency's (DLR) 11$^{\textrm{th}}$ Parabolic Flight Campaign. Microgravity is necessary to investigate low collision velocities that would be inaccessible in the laboratory due to gravity induced sedimentation. A parabolic flight gives around 22 seconds of weightlessness per parabola with a residual acceleration of a few times $0.01\:g_0$ where $g_0$ is acceleration due to gravity on Earth. The flight environment also places some restrictions on one's experiment, with considerations beyond those in a normal laboratory setting. Given that no cryogenic liquids were allowed in the plane, it was necessary to remove the experiment entirely from the aircraft each evening, load the ice particles, pump and cool the system in the laboratory, before re-installing the equipment in the plane just prior to flight. The engineering workarounds to this situation are highlighted below. With limited time and access opportunities to flight, beyond our control, some modifications were also enforced to the types of icy particles we could include in this study - see Section~\ref{Ice particles}.%

\subsection{Experimental apparatus}
\label{Experimental apparatus}
\begin{figure}
   \centering
   \includegraphics[width=\hsize]{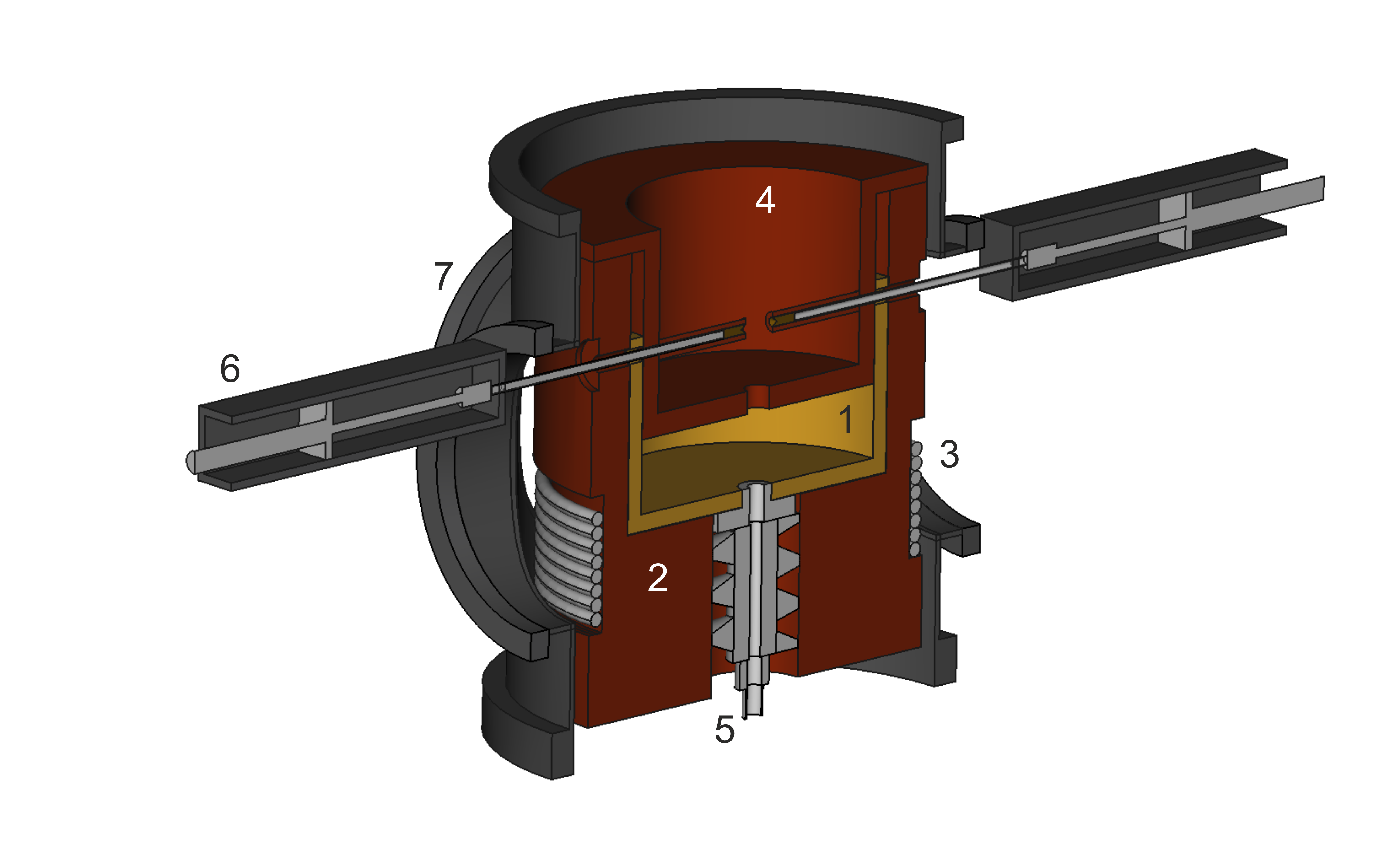}
      \caption{Computer aided design schematic of the experimental setup (adapted from \cite{Salter09}. The particles are stored in a copper colosseum (1) which sits on top of a 45 kg copper block (2) which acts as a thermal reservoir. The copper is passively cooled by passing liquid nitrogen through a feedthrough cooling ring (3) prior to take off. The particles are kept within the colosseum and protected from radiative heating by a copper shield (4). To initiate a collision, the colosseum is rotated (5) to line up two colosseum holes with two diametrically opposed hydraulic pistons (6). The entire set up is enclosed within a vacuum chamber (7).
              }
         \label{FigExp}
   \end{figure}

The experimental apparatus has been previously described in detail \citep{Salter09} and is shown in Fig.~\ref{FigExp}.  Briefly, the experiment comprises two key features; the ability to reach liquid N$_2$ temperatures, i.e. 77 K, and to operate at a residual pressure of around 10$^{-5}$ mBar, thereby negating any residual vacuum water vapour exposure of the particles prior to or during the collisions. Up to 180 particles are stored in a colosseum, a double helix particle reservoir. The colosseum is constructed from copper, which has a high thermal conductivity, and is mounted on top of a 45 kg copper block, which is passively cooled by flowing liquid N$_2$ around a feedthrough cooling ring. The particles are both retained in the colosseum, and further kept from radiative heating by a fully enclosed U-shaped copper heat shield, which is also attached to the copper cooling block. The colosseum moves within this assembly to initiate a collision. The colosseum must be rotated until two of the holes in the colosseum are aligned diametrically opposite each other, and with two accelerating pistons (positioned outside the heat shield and vacuum chamber), as well as the exit guides (attached to the inner heat shield), which lead into the collision volume.  As liquid nitrogen is not permitted on board the parabolic flight aircraft, during flight campaigns the cooling is stopped prior to flight and the copper block then acts as a heat sink (and a cryopump for any water vapour left in the vacuum chamber - leaving the actual vacuum dominated by H$_2$, CO and N$_2$ gas \citep{Fraser02}. Inevitably there is a slow rise in temperature over the course of the flight (see Section~\ref{Effect of temperature}), typically meaning collisions are studied in the range of 130 - 160 K. Since the whole system is encased in a vacuum chamber, this pressure-temperature regime is well below the range at which surface melting, sublimation or desorption of the ice would be anticipated (180 - 200 K) \citep{Ehrenfreund03}.

Once the colosseum is aligned, particle collisions are initiated using the hydraulic piston assembly. The piston heads are attached to the piston assembly outside the vacuum chamber, and feed through a differentially pumped translational flange, so that at rest the head tips retract from the rotating colosseum to sit within the outer region of the U-shaped copper heat shield. The piston tips are made of gold coated copper, providing large thermal conductivity and heat capacity. They are mounted at the end of a hollow stainless steel rod of low thermal conductivity providing the mechanical coupling to the accelerator. Between the experiments the piston heads were retracted in the cold environment of the copper assembly inside the vacuum chamber, being passively cooled by only "seeing" cold walls. Consequently, the piston tips were also cool, and therefore did not induce particle melting during the brief contact between the piston heads and the particles themselves. Using Labview instrument software, and home-designed electronic systems, both pistons were constantly accelerated up to the same velocity, ranging between 0.1 and 0.3~\mss. The pistons then were brought to a hard stop, and rapidly retracted back to their rest position in the U-shaped heat shield. The particles therefore were accelerated out of their storage holes in the colosseum and traveled along (cold) guide tubes to the collision volume with a constant velocity; typically reaching relative velocities of between 0.26 and 0.51 \mss. The entire collision volume was around 2.5 cm $\times$ 2.5 cm $\times$ 2.5 cm. This volume was lit by time-coordinated directional strobe lighting and collisions were recorded using a high speed camera with a frame rate of 107 frames per second and continuous recording capability, across a  2.4 $\times$ 2.0 $\times$ 0.5 region at the centre of the physical collision volume. Mirror optics allowed the camera to capture video footage from two views separated by 60 \ds.

After a collision, particles were still within the vacuum, but outside the cryogenically cooled regions. Fortuitously, the 2 g and 1 g phases of the flight profile forced such particles to sediment to the bottom flange of the vacuum chamber, which was at ambient temperature, and thermally isolated from the passive cooling block.  Consequently, here the particles were rapidly sublimated and the excess water vapour either pumped away, or cryogenically sorbed onto the lower surface of the copper cooling block, thereby never recirculating to the collision volume and certainly unable to reach the other stored particles contained within the colosseum, under the U-shaped heat shield. During flight, after the start of the next parabola, the camera and strobes were started, the colosseum would be rotated to align the next set of stored particles, and the pistons then fired; repeating the whole collision process.

Due to these procedures we are therefore reassured that over the four-hour duration of a flight, our icy particles did not accrete any additional water vapour to their surfaces (forming first a crystalline then amorphous vapour deposited layer sometimes described as frost), nor sublimate such that additional surface roughness was acquired; the particles were stored in a cold (\textless 150 K), shrouded colosseum, in an H$_2$, CO and N$_2$ dominated vacuum at 10$^{-5}$ mBar, thereby negating exposure to water vapour and maintaining the pressure-temperature regime well below any sublimation limits. We return to this point in Section~\ref{Effect of temperature} where we look at the effect of temperature on the results - as temperature increases with time, any effect of frost formation during the course of the experiment would manifest itself as a temperature effect.

\subsection{Ice particles}
\label{Ice particles}
\begin{figure}
   \centering
   \includegraphics[width=\hsize]{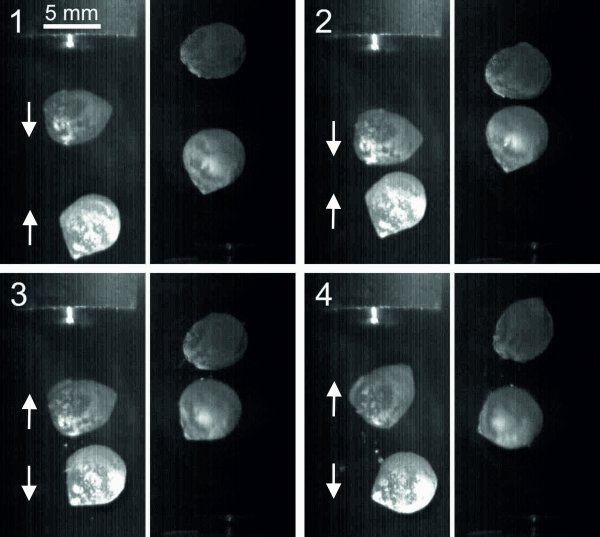}
      \caption{Image sequence of two ice spheres colliding at a relative velocity of 0.42~\ms. The images were captured using mirror optics with the view on the left separated from the view shown on the right by 60\ds. The four successively numbered images were taken 1/107 s apart. The two views appear to be offset due to the set up of the mirror optics.
              }
         \label{FigSpheres}
   \end{figure}

Two types of ice particle were used in this experiment. The first type of particle was small ice spheres, around 5 mm in diameter (see Fig.~\ref{FigSpheres} for in-flight examples). The spheres were prepared by syringing water droplets into liquid nitrogen. The  droplets were formed under the liquid N$_2$ surface, which was then allowed to quiesce until the icy particle and liquid were in equilibrium before forming the next particle. It is known that cooling water droplets of this size using this method will produce ice in its hexagonal phase, as hyperquenched glassy water (a form of amorphous ice) will only be produced if the particles are less than a few microns in diameter \citep{Mayer82, Hallbrucker89, Angell04}.

\begin{figure}
   \centering
   \includegraphics[width=\hsize]{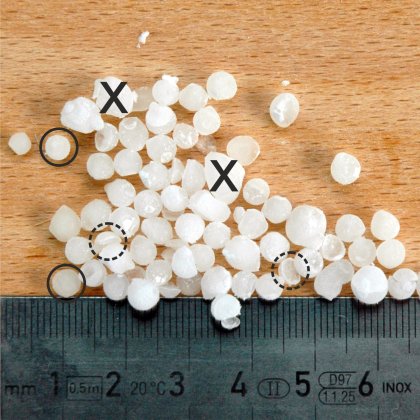}
      \caption{Examples of ice particles produced by syringing water droplets into liquid nitrogen. The polycrystalline nature of the ice is evident from the physical appearance of the particles, as is the anisotropy and roughness of the surfaces. The particles indicated by a solid circle are examples of the sort of particle that was selected for the experiments. Particles such as the ones indicated by a cross were rejected because of their irregular shape and particles such as the ones indicated by a dashed circle were rejected due to their hemispherical shape.
              }
         \label{FigSpheres2}
   \end{figure}

\begin{figure}
   \centering
   \includegraphics[width=\hsize]{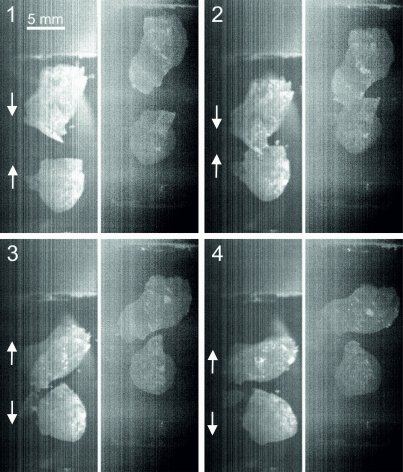}
      \caption{Image sequence of two ice fragments colliding at a relative velocity of 0.36~\ms. The images were captured using mirror optics with the view on the left separated from the view shown on the right by 60\ds. The four successively numbered images were taken 1/107 s apart. The two views appear to be offset due to the set up of the mirror optics.
      }
         \label{FigFrag}
   \end{figure}

 Since such freezing mechanisms are also surface induced, they produce polycrystalline ice (which by its very nature has a cloudy appearance, due to the significant fraction of grain boundaries within the ice sample) with anisotropic surfaces, and which can certainly be described as rough on the molecular level. This can be seen in Fig.~\ref{FigSpheres2}, which shows a sample of ice particles produced by this method, tipped out of their liquid N$_2$ bath into air. It is clear that the polycrystalline ice particles are never transparent (see the examples marked with a solid circle). Of course, unlike the actual ice particles used in this experiment, the ice particles shown in Fig.~\ref{FigSpheres2} have also been exposed to water-vapour from the air for a significant time period, whilst they equilibrated, after outgassing their surface liquid N$_2$ (otherwise we would not get this picture), and therefore a number of them (examples highlighted with an X), have accreted addition surface water vapour (frost). When illuminated equally from all directions by natural light (as opposed to directional strobe lighting), this frost shows up as the non-uniform, high albedo, white regions on these particles. Only those particles of spherical appearance (similar to those marked with a circle in Fig.~\ref{FigSpheres2}) were selected for the experiments; particles that were irregular or hemispherical in shape (such as those outlined with a dotted line) were not chosen. It should be noted that when the particles are illuminated by directional strobe lighting in flight, the combination of directional lighting and rough surfaces produces areas of high albedo (see Figs.~\ref{FigSpheres} and \ref{FigFrag}) which should not be mistaken for frost.

 The second type of ice particle was created by submerging a spoonful of water in liquid nitrogen and then crushing the resulting ice with a hammer to create irregularly shaped ice fragments (see Fig.~\ref{FigFrag} for in-flight examples). The fragments ranged in size from 4.7 to 10.8 mm in diameter (diameter here is the length of the major axis of an ellipse fitted to the images).  Again, all particles were stored under liquid nitrogen until they were individually loaded into the particle reservoir which had been pre cooled to 77 K, using the method described above. For the first day of the campaign, the ice spheres were used. However, due to the spherical nature of the particles, a large number of them rolled out of their holders during flight. This meant that few good collisions were obtained and so for the remaining two days of the campaign, the ice fragments were used, to maximise the data collected in the limited flight time available. This also gives us the unique opportunity to study the collisional properties of irregularly shaped ice fragments which are likely to bear a closer resemblance to icy particles found in planet forming regions than spherical samples, as particles in protoplanetary disks are assumed to be irregular in shape \citep{Mutschke09, Perry12, Min12}.

Particles were selected for our experiments from those prepared and stored under liquid N$_2$. Sample loading was conducted entirely in the laboratory, just before a flight. Using tweezers (which had also been cooled and equilibrated in liquid N$_2$), particles were selected one by one from the liquid N$_2$ bath, stood at the centre of the collision volume, a few cm from the pre-cooled colosseum. The particles, with a nominal temperature of 77 K, were then transferred directly to a hole in the colosseum, also cooled to 77 K, and exposed by removing the inner part of the U-shaped heat shield. Within a few minutes, whilst particles were still outgassing N$_2$, the U shield was reattached, and the whole system, whilst still cooling at 77 K, evacuated to 10$^{-5}$ mBar. With significant cold sorbing surfaces in the vicinity of the particle loading operation, as well as the rapid evaporation of liquid N$_2$ from the particles as they are loaded, we attempt to minimise any water vapour deposition (frost growth) on to our particles during this step. Such a process might occur on the timescales of minutes that this procedure takes, depending on the ambient laboratory humidity and temperature. Of course, in comparison to the timescales of flight itself (which we have illustrated above have no measurable frosting or sublimation effects on the particles), this preparation period is very short, but nonetheless should not be neglected. The geometry of our experiment dictated that the first particles loaded to the colosseum would be the last to be collided in flight, so as a function of time (indicated by temperature, as the temperature increases over the course of a flight), \textit{if} the loading process did lead to further surface coating of the ice spheres, one might expect to see this reflected in a clear trend in the resulting coefficient of restitution data with temperature (time) (see Section~\ref{Effect of temperature}).

\subsection{Collision statistics}
\label{Collision statistics}

A total of ninety-three parabolas were available during this campaign. With multiple collisions possible per parabola, there was the potential for a larger number of collisions. However, not all the collisions were successful, and of those that were, not all were suitable for analysis. A first count of the collisions gave 104 successful collisions. Of these, 52 were unsuitable for analysis for reasons detailed in  Table~\ref{collisiontable}, leaving us with 52 suitable for the quantitative analysis which follows.

\begin{table}
\caption{Clarification of collisional outcomes. The binary collisions suitable for analysis are focussed on in the resulting analysis while the fragmentation and rotation before the collision events are discussed separately.}             
\label{collisiontable}      
\centering                          
\begin{tabular}{c c}        
\hline\hline                 
Type of collision & Number of collisions \\    
\hline                        
   Binary collisions suitable for analysis & 52  \\      
   Non binary & 33  \\
   Poor image quality & 7\\
   Residual acceleration & 4\\
   Multiple hit & 4\\
   Fragmentation & 4  \\
   \hline
   Total & 104 \\

\hline                                   
\end{tabular}
\end{table}

\begin{figure}
   \centering
   \includegraphics[width=\hsize]{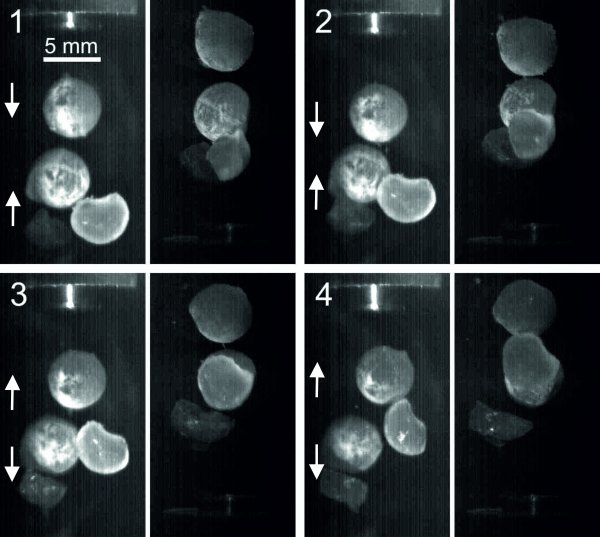}
      \caption{Image sequence of four ice particles colliding. The images were captured using mirror optics with the view on the left separated from the view shown on the right by 60\ds. The four successively numbered images were taken 1/107 s apart. The two views appear to be offset due to the set up of the mirror optics.
              }
         \label{FigNonBinary}
   \end{figure}

\begin{figure}
   \centering
   \includegraphics[width=\hsize]{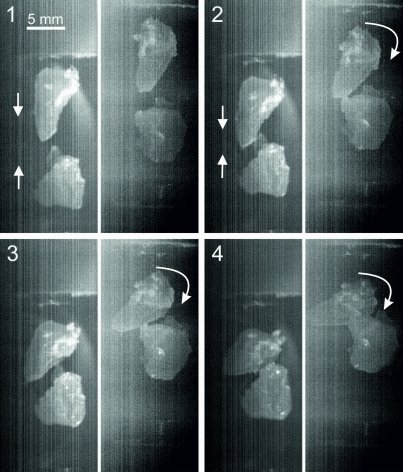}
      \caption{Image sequence of two ice fragments colliding at a relative velocity of 0.34~\ms and hitting multiple times. The curved arrows shows the direction of rotation. Images after the collision are not shown because both particles hit the top piston before they separated. The images were captured using mirror optics with the view on the left separated from the view shown on the right by 60\ds. The four successively numbered images were taken 1/107 s apart. The two views appear to be offset due to the set up of the mirror optics.
      }
         \label{FigMultipleHit}
   \end{figure}

Non binary collisions (Fig.~\ref{FigNonBinary}) were excluded from the analysis as the moment of collision was rarely captured, making it impossible to determine whether these were truly multiple body collisions or repeated two body collisions. It is likely that multiple body collisions occur in protoplanetary disks and so considering this type of event would give further information, however this is not possible to do with the current data.
Poor image quality hampered analysis efforts in several cases. In a few cases, residual acceleration of the plane caused the particles to move with respect to the camera, meaning they were no longer in the field of view.
Multiple hits of the two particles in quick succession occurred in some cases (Fig.~\ref{FigMultipleHit}) which made it impossible to determine how the velocity changed for each hit. These types of event were only seen for fragment collisions where there was rotation after the collision and the shape of the particles was such that rotation caused a multiple hit. As the particles in protoplanetary disks are likely to be irregular in shape, it is likely that rotation will cause multiple hits in quick succession. The effects of rotation are discussed later in the paper.
Fragmentation events were removed from the main analysis due to the difficulties in tracking multiple fragments after collision but are discussed separately later in the paper.
Six binary collisions of ice spheres and 46 binary collisions of ice fragments suitable for analysis were captured. The results of the analysis are presented in Section~\ref{Results and discussion}.

\section{Analysis methodology}
\label{Analysis methodology}
The video images were analysed and collision parameters were extracted in the following way. The entire flight footage was viewed and a record was kept of all collisions. Each individual collision was extracted and viewed frame by frame to check that the collision was real and suitable for analysis. Beam splitter optics allowed two different views of each collision to be captured separated by an angle of 60\ds. The particles were manually tracked from the image sequences in both views and the positions obtained were converted to three dimensional co-ordinates using a transformation algorithm. Performing linear fits to these co-ordinates yielded particle trajectories which then gave relative velocities before and after the collision. The distance of closest approach of the particles could also be extracted from the particle trajectories; using this and the relative collision velocity yielded the impact parameter. Coefficients of restitution were calculated from the relative velocity before and after the collision, and were also resolved into components tangential and normal to the colliding surfaces, giving further information about the collision.

Error analysis was performed on all the calculations for relative impact velocity and coefficient of restitution using standard methods for calculating the propagation of errors. For the normalised impact parameter, an average error is given taking into account the geometry of the particles and measurement errors.

\section{Results and discussion}
\label{Results and discussion}

\subsection{Coefficient of restitution and impact parameter}
\label{Coefficient of restitution and impact parameter}
 The dominant collisional outcome was bouncing. Fragmentation was observed in a few cases but in no case did the ice particles stick. If solid ice particles have a "bouncing barrier" as dust aggregates do \citep{guettleretal2010,Zsom10,weidlingetal2012,kotheetal2013}, it would appear that we are in this region. Hence the coefficient of restitution can be calculated to give an indication of how much translational kinetic energy was lost in each collision.

\begin{figure}
   \centering
      \includegraphics[width=\hsize]{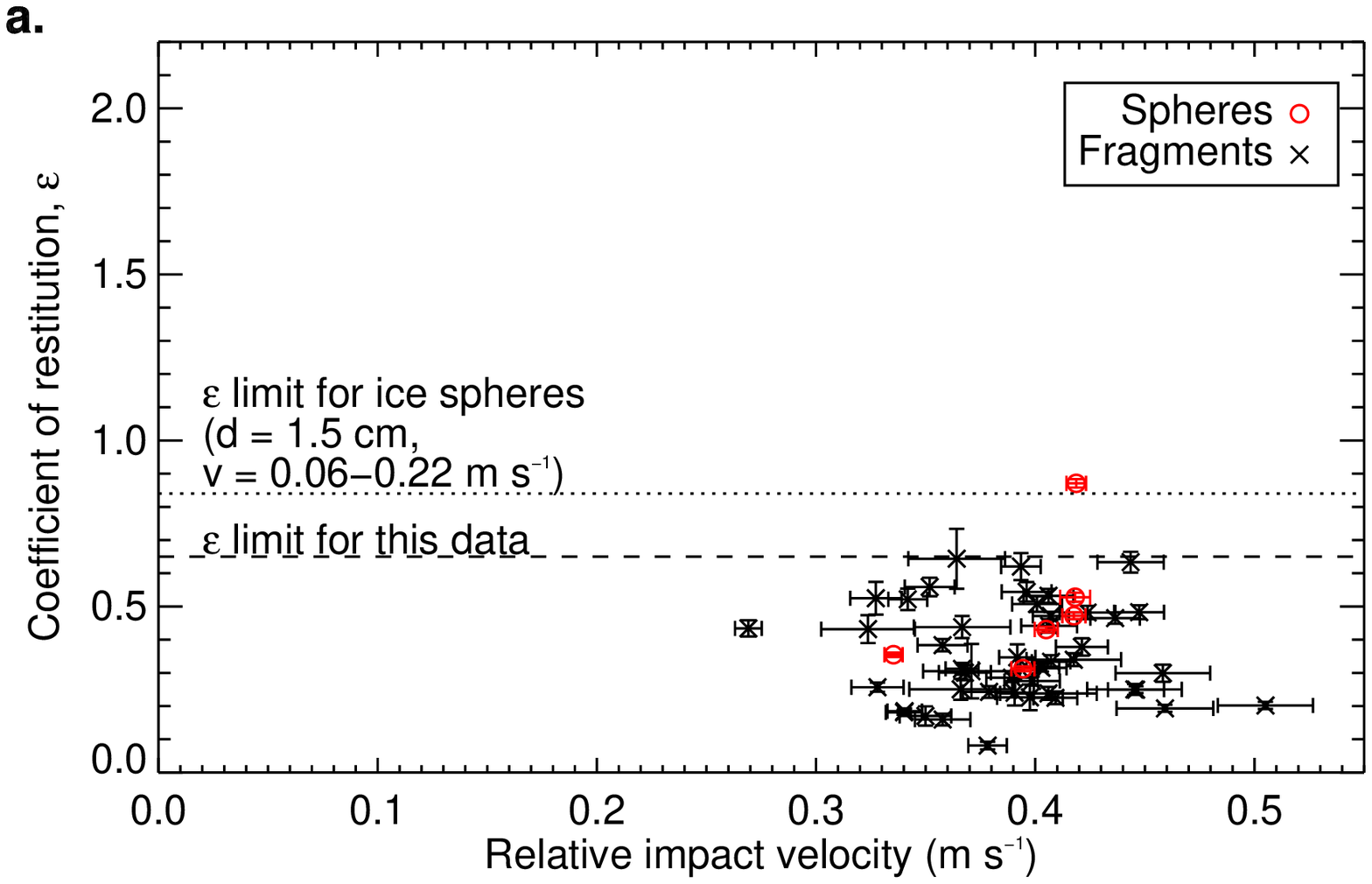}
      \includegraphics[width=\hsize]{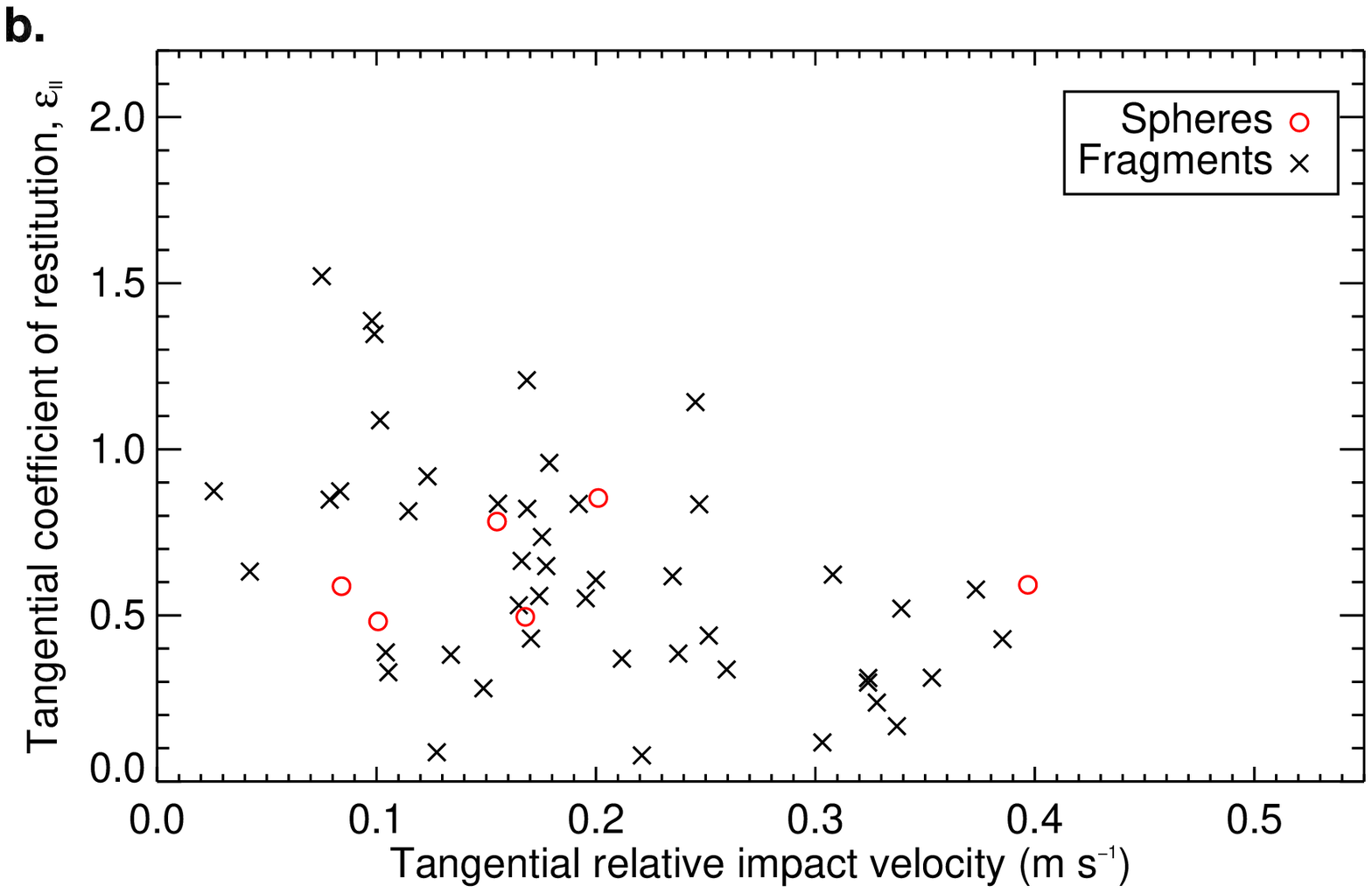}
      \includegraphics[width=\hsize]{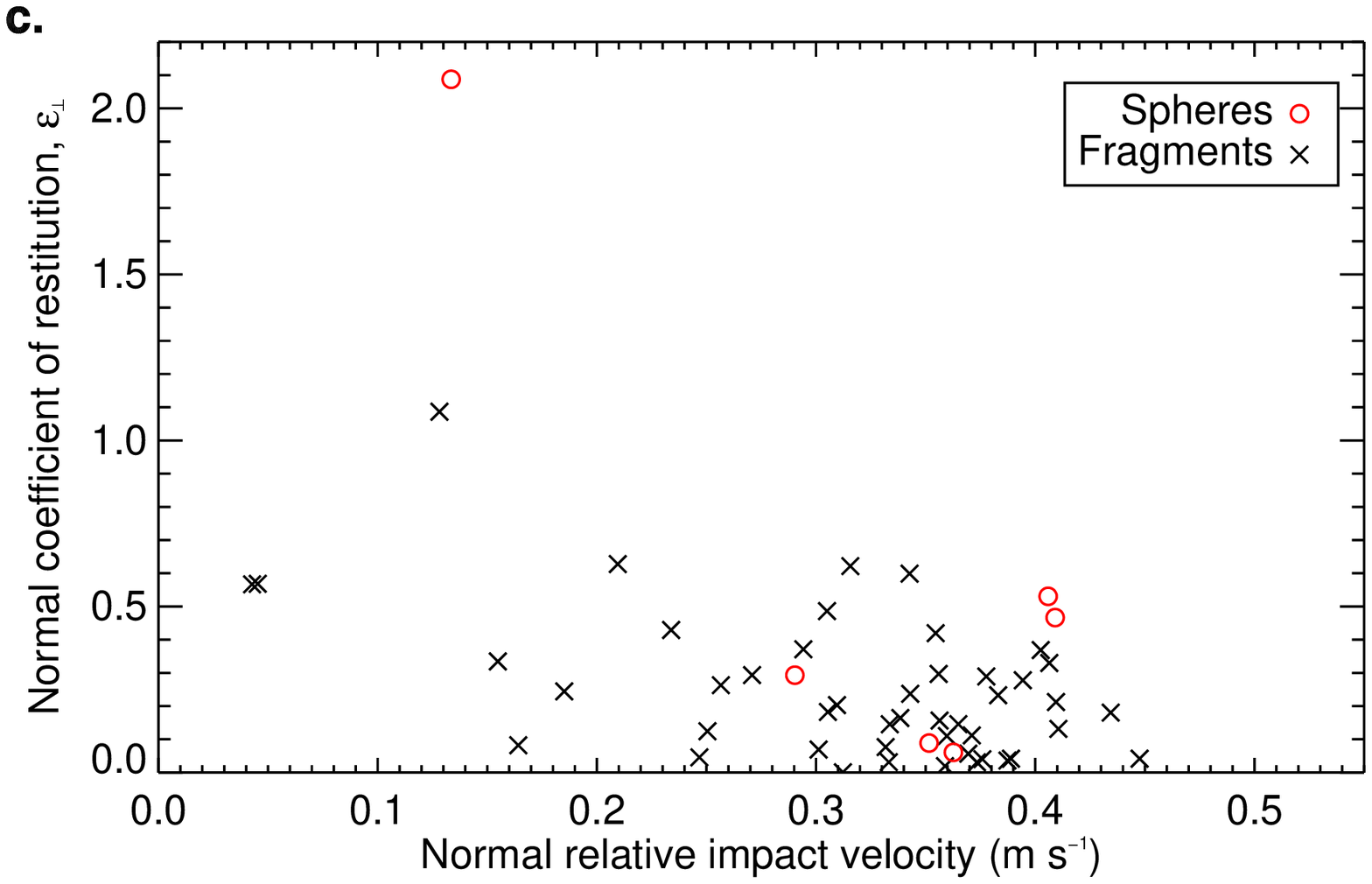}
      \caption{Coefficient of restitution as a function of relative impact velocity. \textbf{a.} Relative impact velocity versus coefficient of restitution. The dashed lines show the limits of coefficient of restitution for this data and for collisions of 1.5 cm ice spheres in the velocity range 0.06-0.22 \ms (found by \cite{Heisselmann10}). \textbf{b.} Relative impact velocity versus coefficient of restitution tangential to the colliding surfaces. \textbf{c.} Relative impact velocity versus coefficient of restitution normal to the colliding surfaces.
      }
         \label{Figrvball}
   \end{figure}

\begin{figure}
   \centering
   \includegraphics[width=\hsize]{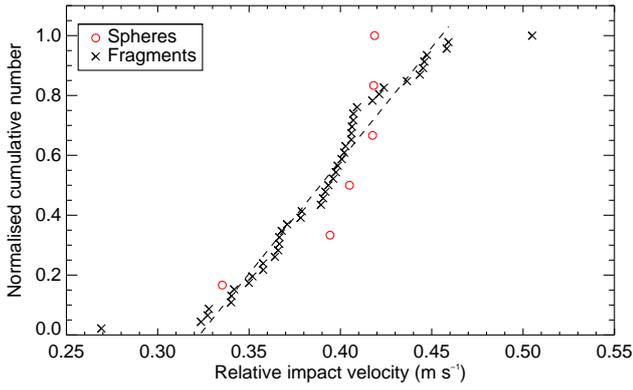}
      \caption{Normalised cumulative number for collisions with relative impact velocity $\leq v$. The data has been fitted with a linear fit between 0.32 and 0.46~\ms for the fragments (dashed line). A good straight line fit was not obtained for the spheres but due to the limited number of data points, this is not a cause for concern.
              }
         \label{Figvcumul}
   \end{figure}

\begin{figure}
   \centering
   \includegraphics[width=\hsize]{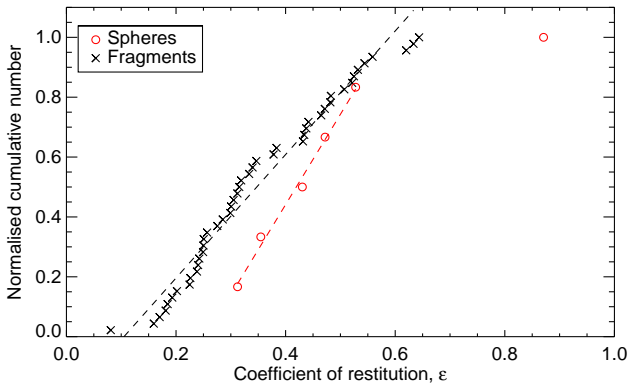}
      \caption{Normalised cumulative number for collisions with coefficient of restitution $\leq \varepsilon$. The data has been fitted with a linear fit across the entire range for the fragments and between 0.31 and 0.53 for the spheres (dashed lines).
              }
         \label{FigCORcumul}
   \end{figure}

\begin{figure}
   \centering
   \includegraphics[width=\hsize]{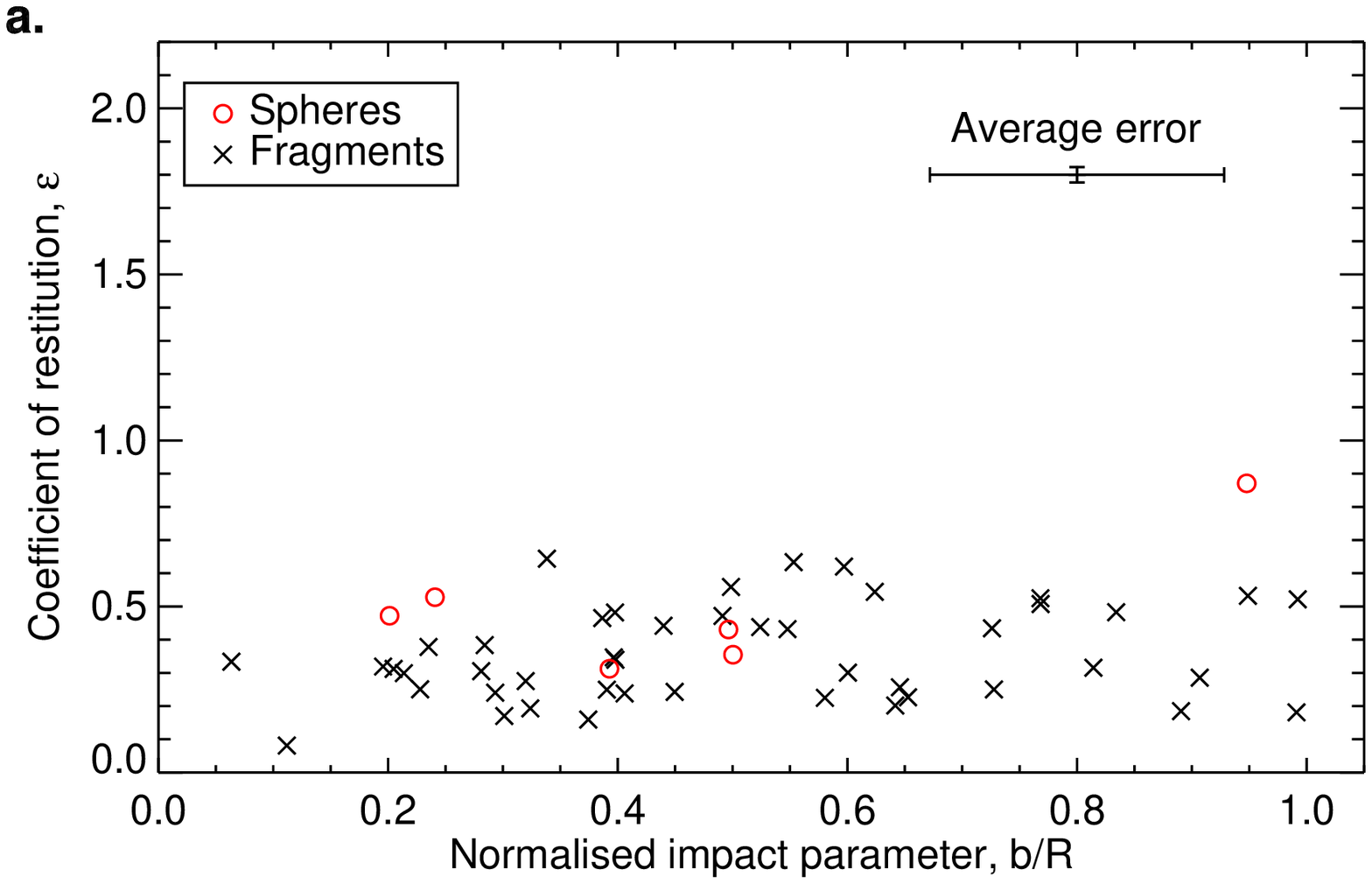}
   \includegraphics[width=\hsize]{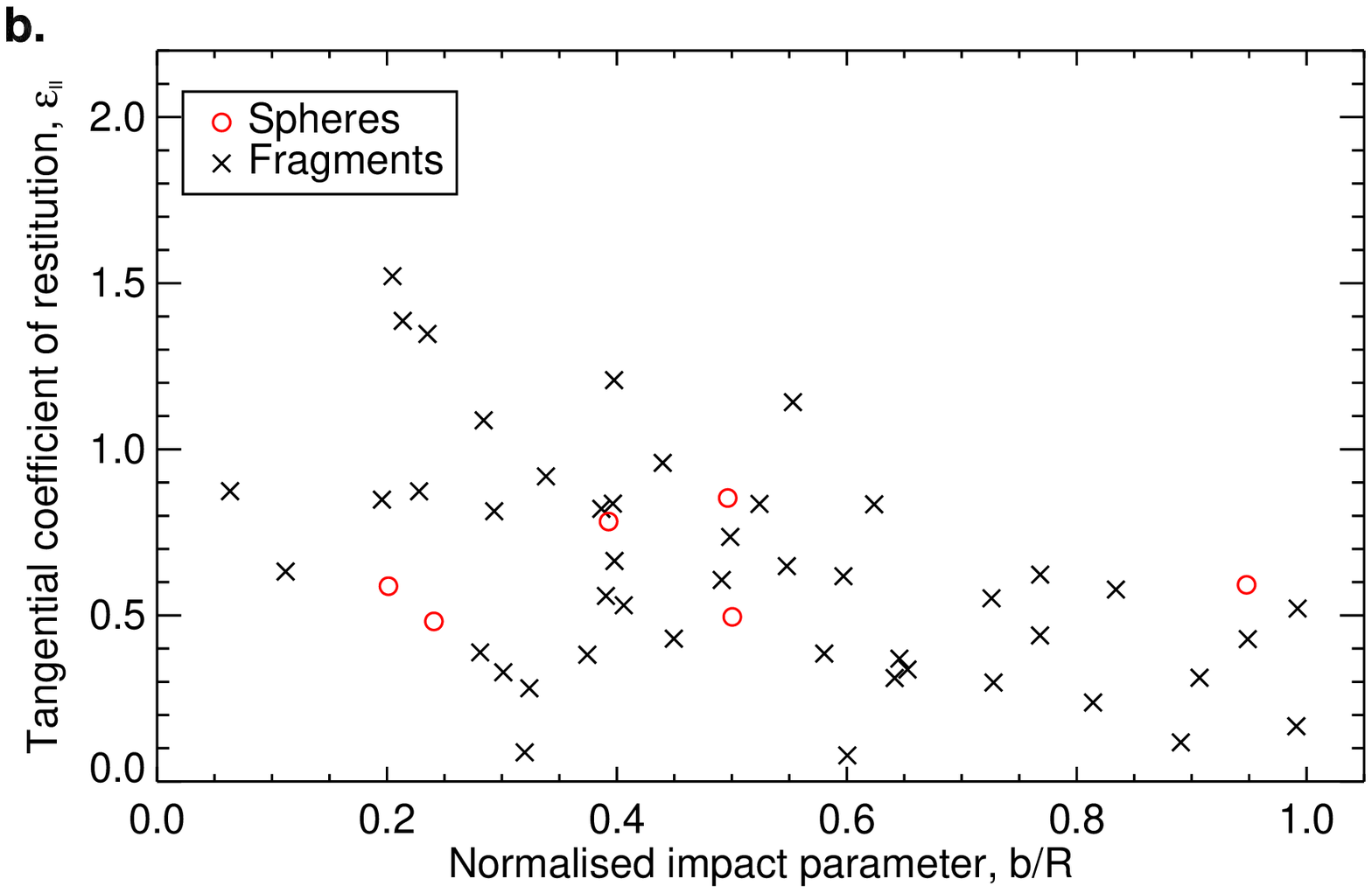}
   \includegraphics[width=\hsize]{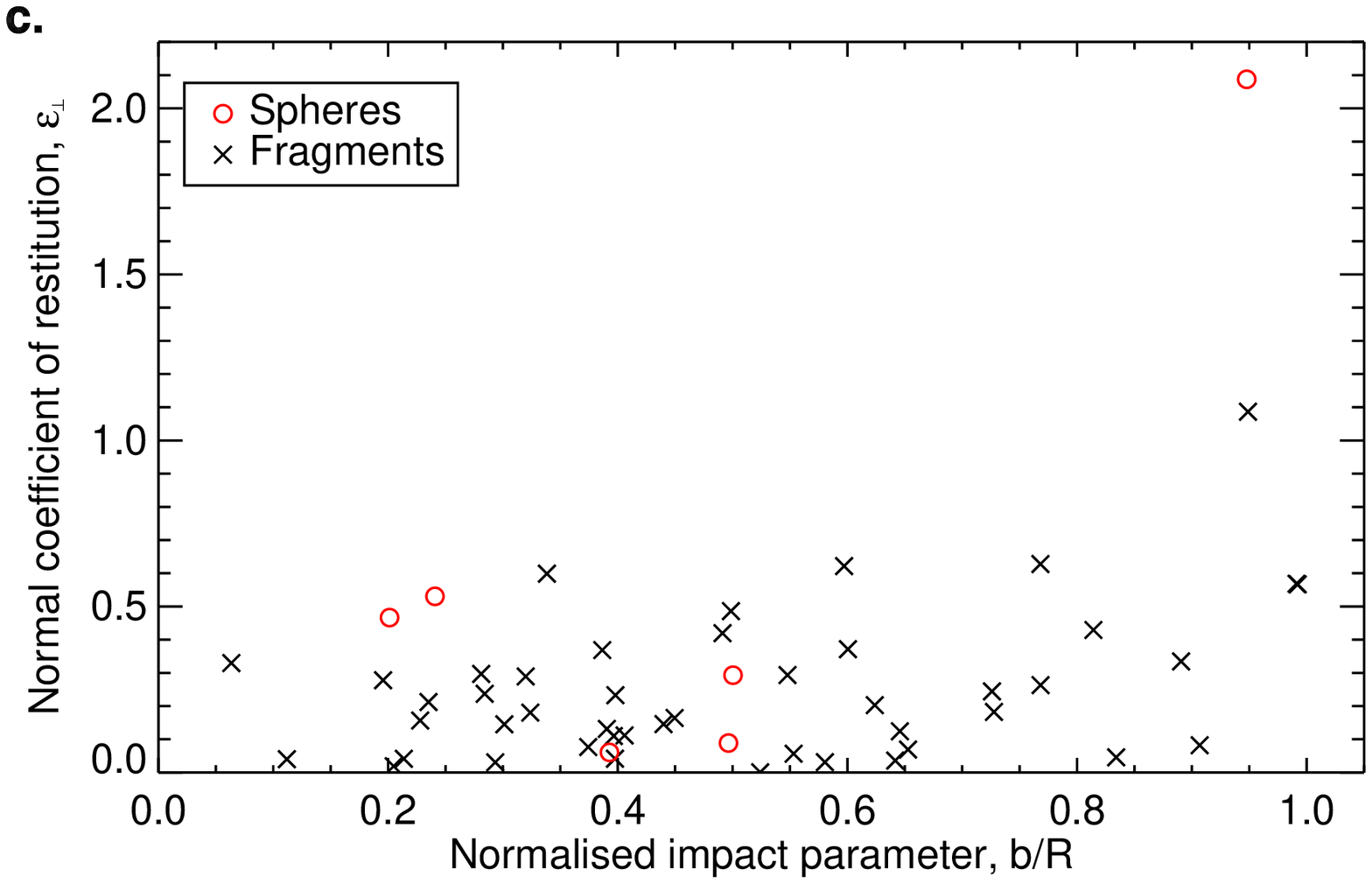}
      \caption{Coefficient of restitution as a function of normalised impact parameter. \textbf{a.} Obtained normalised impact parameter versus coefficient of restitution. \textbf{b.} Obtained normalised impact parameter versus coefficient of restitution tangential to the colliding surfaces. \textbf{c.} Obtained normalised impact parameter versus coefficient of restitution normal to the colliding surfaces.
      }
         \label{Figipall}
   \end{figure}

\begin{figure}
   \centering
   \includegraphics[width=\hsize]{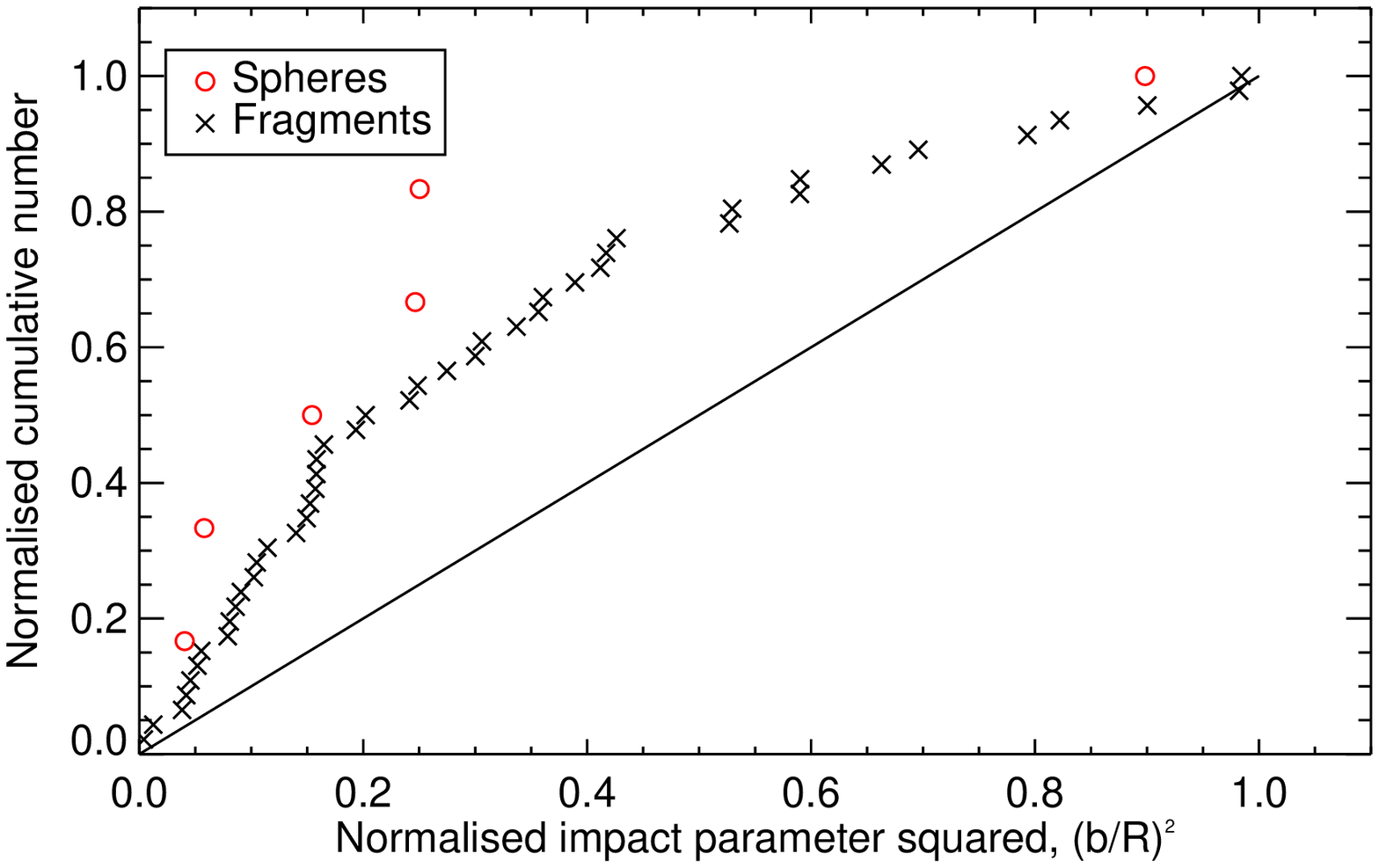}
      \caption{Normalised cumulative number for collisions with  squared normalised impact parameter $\leq $ (\textit{b/R})$^2$. The solid line indicates the expected distribution for perfectly random collisions. The spread demonstrates that our apparatus slightly favours head on collisions over glancing ones.
              }
         \label{Figipsqcumul}
   \end{figure}

Fig.~\ref{Figrvball} shows the spread of coefficient of restitution with relative impact velocity. The impact velocities are distributed from 0.27 to 0.51~\ms and are evenly distributed between  0.32 and 0.46~\ms, which is shown by the linear fit for the fragments in the range 0.32 and 0.46~\ms (Fig.~\ref{Figvcumul}). A good linear fit was not obtained for the spheres but due to the limited number of data points this is not a cause for concern.
Coefficients of restitution are spread evenly between 0.08 and 0.65 (Fig.~\ref{FigCORcumul}) apart from one outlier with a coefficient of restitution of 0.87.  The average coefficient of restitution (ignoring the outlier) is 0.36.

To test the correlation between the coefficient of restitution and impact velocity the linear Pearson correlation coefficient  was computed: \begin{equation}r=\frac{\sum_{i=0}^n (y_i-\bar{y})(x_i-\bar{x})}{\sqrt{\sum_{i=0}^n (y_i-\bar{y})^2}\sqrt{\sum_{i=0}^n (x_i-\bar{x})^2}} \end{equation} The number of points is given by \textit{n} and the mean values of \textit{x} and \textit{y} are given by $\bar{x}$ and $\bar{y}$ respectively. This gives a measure of the strength of correlation between two variables, \textit{x} and \textit{y}, with a value of 1/-1 indicating perfect positive/negative correlation and a value of 0 indicating no correlation. The correlation between the coefficient of restitution and the relative impact velocity was -0.05 for the fragments and 0.53 for the spheres. It is clear that there is no correlation for the ice fragments, or for the ice spheres; the limited number of data points for the spheres means that a value of 0.53 cannot be said to indicate high correlation.

The range of coefficients of restitution instead of one single value, as well as the lack of correlation with impact velocity are results that were also found by \cite{Heisselmann10}, and do not replicate the findings of earlier results \citep{Hatzes88, Higa96, Higa98}. A comparison of our experimental methods with those of previous work is shown in Table~\ref{literaturetable}. It can be seen that along with the work of \cite{Heisselmann10}, our study is unique in both the method of collision (using pistons to collide same size particles under microgravity conditions) and in the pressure-temperature regime used (130-180 K and $10^{-5}$ mbar). This pressure-temperature regime is vital for ensuring that frost does not form on the particle surfaces (see Section~\ref{Experimental apparatus}). Also, the surfaces of our particles are not smooth like the particles in the work of \cite{Hatzes88}, \cite{Higa96} and \cite{Higa98}. The range of coefficients of restitution is likely due to the rough and anisotropic nature of our particle surfaces, in contrast to previous work where care was taken to ensure smooth surfaces. It is also possible that both the method of collision (which is arguably a more realistic analogue for collisions happening in protoplanetary disks) and the use of same sized ice particles rather than an ice particle and a target to determine coefficients of restitution also play a role.

\begin{table*}
\caption{A comparison of the particle sizes and experimental conditions in ice collision studies to date. The particle sizes, collision methods, surface properties,collision velocities, temperature and pressure are all shown for comparison. }            
\label{literaturetable}      
\centering                          
\begin{tabular}{c c c c c c c c}        
\hline\hline                 
Study & Size of particle(s) & Collision method & Surface  & Relative collision & Temperature & Pressure\\
  &  (diameter) &  & properties & velocities (\mss) & (K)  &  (mbar)\\   
\hline                        
 Hatzes et al.  & 50 mm ice sphere  & Pendulum & Smooth, roughened  & 0.00015-0.02  & 85-145  & $1.3 \times 10^{-3}-$  \\      
  (1988)& and ice block & & and frosted & & & $1.3 \times 10^{-4}$ \\
Higa et al.   & 30 mm ice sphere  & Free fall & Smooth & 0.01- 7 & 113-269  & 10\\
 (1996 & and ice block & & & & & \\
 Higa et al. & 2.8- 72 mm & Free fall & Smooth & 0.01-10  & 261  & 1000\\
 (1998) & sphere and  & & & & & \\
  & ice block & & & & & \\
 Hei{\ss}elmann & 15 mm & Pistons, & Rough, unfrosted & 0.06-0.22 & 130-180  & Not stated   \\
  et al. (2010)& spheres & microgravity& & & & but likely to  \\
    & & & & & &  be around \\
   & & & & & &  $10^{-5}$  \\
 Present work  & 4.7-10.8 mm & Pistons, & Rough, unfrosted & 0.26-0.51 & 130-160  & $10^{-5}$ \\
  & spheres and & microgravity& & & &  \\
    & fragments & & & & &  \\

\hline                                   
\end{tabular}
\end{table*}

Fig.~\ref{Figipall} shows the spread of coefficient of restitution with the normalised impact parameter \textit{b/R}. The distribution of the squared impact parameter is shown in Fig.~\ref{Figipsqcumul}. Perfectly random collisions should follow a straight line between the points (0,0) and (1,1) in three-dimensional space (solid line in Fig.~\ref{Figipsqcumul}). As can be seen, our apparatus slightly favours collisions with impact parameters closer to the head-on collisions over grazing incidences. The linear Pearson correlation coefficient was again computed for the correlation between coefficient of restitution and normalised impact parameter, yielding a value of 0.23 for the fragments and 0.69 for the spheres. Once again, there is no correlation for either the fragments or the spheres. It can be seen from Fig.~\ref{Figipall}.a. that the collision with a coefficient of restitution of 0.87 was a very glancing one with a very high impact parameter, which explains the smaller loss of energy in this case. Hence we are justified in stating the 0.65 limit for the coefficient of restitution in the majority of cases.

As indicated in Fig.~\ref{Figrvball}.a., the coefficients of restitution have an upper limit of 0.65. Similar experiments by \cite{Heisselmann10} on collisions of 1.5 cm (diameter) ice spheres in the velocity range 0.06-0.22 \ms gave an upper limit of 0.84 (also shown in Fig.~\ref{Figrvball}.a.). This demonstrates that the limit of coefficient of restitution decreases with increasing impact velocity, which is corroborated by \cite{guettleretal2012} and \cite{krijtetal2013}. It is also possible that this is a result of deviation from sphericity (compared to the results of \cite{Heisselmann10} which used almost perfectly spherical samples) or due to the smaller size of our particles.

Fig.~\ref{Figrvball}.b. and c. and Fig.~\ref{Figipall}.b. and c. show the coefficient of restitution tangential ($\varepsilon_\parallel$) and normal ($\varepsilon_\perp$) to the colliding surface versus the tangential and normal impact velocity  and the normalised impact parameter respectively. This was calculated by resolving the impact and rebound velocities into tangential and normal components and calculating the tangential and normal coefficients of restitution from these: \begin{equation} \varepsilon_\parallel = \frac{v_{a\parallel}}{v_{b\parallel}} \end{equation} \begin{equation} \varepsilon_\perp = \frac{v_{a\perp}}{v_{b\perp}} \end{equation} where \textit{a} and \textit{b} denote after and before the collision respectively. Resolving the coefficient of restitution into these components gives information about the distribution of translational kinetic energy into rebound (normal) and scattering (tangential). Where values are above 1 this indicates that the corresponding velocity component after the collision was greater than the component before. This is possible due to velocity transfer between components.  The values for $\varepsilon_\parallel$ are distributed from 0 to 2.1 and show no limit at 0.65 (Fig.~\ref{Figrvball}.b., Fig.~\ref{Figipall}.b.). Considering simple geometry, a lower value of normalised impact parameter indicates a lower tangential component to the relative impact velocity. Apart from two cases, the values for $\varepsilon_\perp$ are distributed between 0 and 0.65 similarly to the overall values of $\varepsilon$. From Fig.~\ref{Figipall}.c. it can be seen that the cases where $\varepsilon_\perp$ is greater than 1 are very glancing collisions (\textit{b/R} = 0.95) so will have a proportionally smaller component of relative impact velocity perpendicular to the colliding surfaces. The two collisions with very low normal impact velocities both have very high normalised impact parameters (both 0.99) which is to be expected, as a consideration of the geometry indicates that a glancing collision will have a low velocity component normal to the colliding surfaces. There is a slight negative correlation between $\varepsilon_\perp$ and normal relative impact velocity (correlation coefficient of -0.51 for fragments). There is no correlation (correlation coefficient of 0.40 for fragments) between $\varepsilon_\perp$ and normalised impact parameter. From the collision footage it can be seen that the particles mainly rebound rather than scatter and so the overall value for $\varepsilon$ is likely to be dominated by the perpendicular component; hence the limit at 0.65 is seen for $\varepsilon_\perp$ but not for $\varepsilon_\parallel$.

These results show that a minimum of 58\% (1-$\varepsilon^2$ for $\varepsilon = 0.65$) of the translational kinetic energy is lost in collision, in some cases more.  In most cases it can be seen that some energy is converted into rotational energy which is the subject of the next section.
\subsection{Rotation}
\label{Rotation}
\begin{table}[h]
\caption{The percentage of particles in the binary collisions that were suitable for analysis that rotate and do not rotate before and after the collision.}             
\label{rotationtable}      
\centering                          
\begin{tabular}{c c c c}        
\hline\hline                 
      & Rotates (\%) & Does not rotate (\%) & Unclear (\%) \\    
\hline                        
   Before & 10 & 89 & 1 \\
   After & 84 & 6	& 10 \\
\hline                                   
\end{tabular}
\end{table}

\begin{table*}[ht]
\caption{Energy before and after the collision as a percentage of the total translational and rotational kinetic energy before the collision. The coefficient of restitution is also shown.}             
\label{energytable}      
\centering                          
\begin{tabular}{c c c c c c c}        
\hline\hline                 
 Case  &	Translational  & Rotational   &	Translational  & Rotational  &	Unaccounted   & Coefficient\\
  number & energy before (\%) & energy before (\%) & energy after (\%)  & energy after (\%) & for energy after (\%) & of restitution\\
\hline                        
1&  100.00 & 0.00 &  8.14&	17.17&	74.68 & 0.29\\
2&	94.02 & 5.98 & 26.60&	1.70&	71.70 & 0.53\\
3&	100.00 & 0.00 &  9.95&	3.33&	86.72 & 0.32\\
4&	100.00 & 0.00 &  2.54&	1.55&	95.91 & 0.16\\
5&	100.00 & 0.00 &  5.12&	1.31&	93.57 & 0.23\\
6&	100.00 & 0.00 &  3.73&	0.08&	96.18 & 0.19\\
7&	100.00 & 0.00 &  31.22&	3.80&	64.98 & 0.56\\
8&	100.00 & 0.00 &  5.78&	2.60&	91.62 & 0.24\\
9&	100.00 & 0.00 &  41.43&	0.20&	58.37 & 0.64\\
10&	100.00 & 0.00 &  19.51&	12.57&	67.92 & 0.44\\

\hline                                   
\end{tabular}
\end{table*}

\begin{figure}[h]
   \centering
   \includegraphics[width=\hsize]{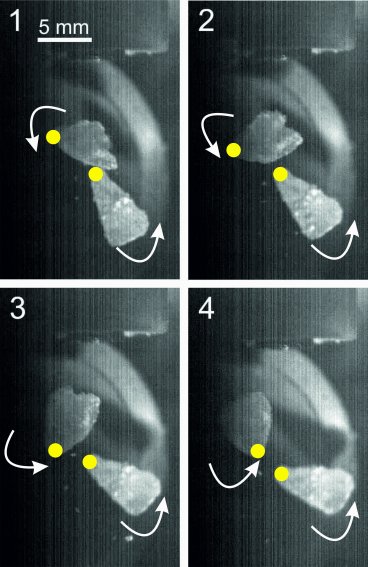}
      \caption{Image sequence of two ice fragments rotating after collision at angular velocities of -28~rad~s$^{-1}$ (top) and -11~rad~s$^{-1}$ (bottom) where rotation in the clockwise direction is defined as positive. The four successively numbered images were taken 2/107 s apart. The spots tracked are marked in yellow and the arrows show the direction of rotation.}
         \label{FigRotation}
   \end{figure}

The percentage of particles in the binary collisions suitable for analysis is shown in Table ~\ref{rotationtable}. The majority of the particles do not rotate before collision (89\%), whereas after the collision the majority do rotate (84\%). All of the particles that rotated before the collision continued to rotate afterwards and in all cases at least one particle was rotating after the collision. It has been demonstrated in numerical simulations that ice particles can lose translational kinetic energy into rotation \citep{Schafer07}. However, to date this has not been studied experimentally in a quantitative way. Quantifying particle rotation presents a considerable experimental challenge; the particle must be marked in some way and this mark must be tracked across subsequent images. The particle must also rotate in the field of view of the camera. We were not able to mark our particles but due to the nature of our ice fragments, it was possible to identify distinguishing marks and track these where they rotated in the camera field of view for 10 cases. An example is shown in Fig.~\ref{FigRotation}. The percentage of energy that went into rotation and translation (considering the translational and rotational energy before the collision) is shown in Table ~\ref{energytable}, with the percentage of energy unaccounted for. The amount of energy that is converted into rotation varies from 0.08\% to 17\%, whereas the energy that is converted into translation varies from 4\% to 41\%. The energy unaccounted for ranges from 58\% to 96\%. These results make it clear that the quantity of energy converted into rotation is not sufficient to account for the loss of translational kinetic energy observed. A study by \cite{Zamankhan10} showed that most translational kinetic energy is dissipated due to surface fracturing. This would be difficult to observe in our results as any such fracturing is likely to be small and not visible in the video images. In some cases, fragmentation was observed (see next section). Previous work on ice collisions has demonstrated that frost (condensed water or other liquid) on particle surfaces can reduce the coefficient of restitution (\cite{Hatzes88}, \cite{Supulver97}); however, due to our experimental procedures detailed in Section~\ref{Ice particles} it is unlikely that frost is present on our surfaces. Other possibilities include compaction and desorption of material from the surface of the particles (which could be facilitated by surface heating induced by friction due to the rough surfaces of the particles or dissipation of collision energy into the surface during the contact between particles), both of which are beyond the scope of this study.

\subsection{Fragmentation}
\label{Fragmentation}

\begin{figure}[h]
   \centering
   \includegraphics[width=\hsize]{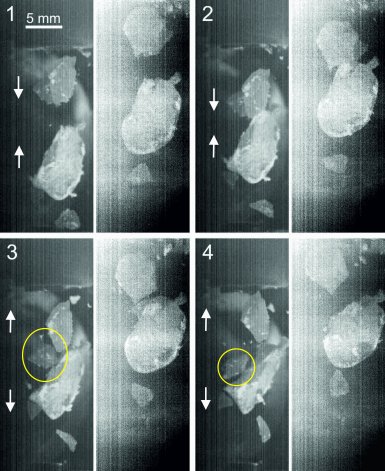}
      \caption{Image sequence of two ice fragments colliding at a relative velocity of 0.42~\ms and fragmenting (fragment shown circled in yellow). The images were captured using mirror optics with the view on the left separated from the view shown on the right by 60\ds. The four successively numbered images were taken 1/107 s apart. The two views appear to be offset due to the set up of the mirror optics.
      }
         \label{FigFragmentation}
   \end{figure}

The collisional outcomes in these experiments were almost universal bouncing. However, for a small number of collisions, fragmentation occurred. In one case the fragmentation was catastrophic (Fig.~\ref{FigFragmentation}), whereas the others only involved a small amount of fragmentation. Previous research on dust aggregation has shown that above a certain velocity, aggregates will fragment \citep{blumwurm2008,beitzetal2011,schraepleretal2012}. This critical velocity has been reported as 1.24~\ms for ice spheres of 2.8~mm diameter and as 0.702~\ms for ice spheres of 8 mm diameter \citep{Higa98}; our particles are around these sizes at between 4.7 and 10.8 mm in diameter and so we would expect the critical velocity for fragmentation to be in the same range. However, the relative impact velocities for our fragmentation events were between 0.34 and 0.42~\ms, which is within the range of impact velocities for the collisions that resulted in bouncing, and certainly considerably less than the critical velocity for fragmentation. It is therefore likely that the fragmentation was a result of prior weakening of the particles, for example fracturing in their creation. Fragmentation can only be observed when the fragments are larger than the observable pixel size of the camera, in this case $39 \times 39$ \textmu m$^2$. Hence it is possible that very small amounts of fragmentation occur in many more collisions.

\subsection{Effect of temperature}
\label{Effect of temperature}

\begin{figure}[h]
   \centering
   \includegraphics[width=\hsize]{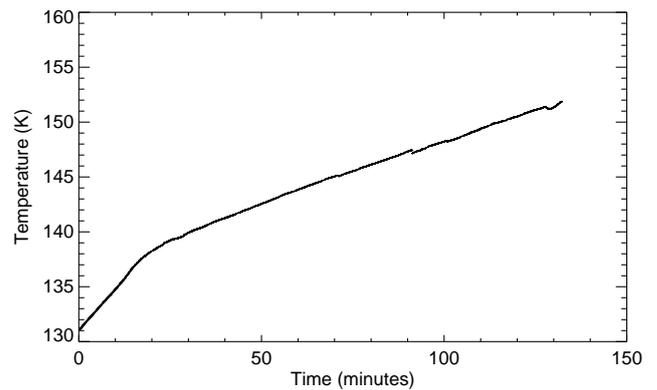}
      \caption{Increase in temperature with flight duration. The temperature increases at a rate of 22.5 K h$^{-1}$ for the first 20 minutes and at a rate of 7.0 K h$^{-1}$ for the remainder of the flight.
      }
         \label{FigTimeTemp}
   \end{figure}

\begin{figure}[h]
   \centering
   \includegraphics[width=\hsize]{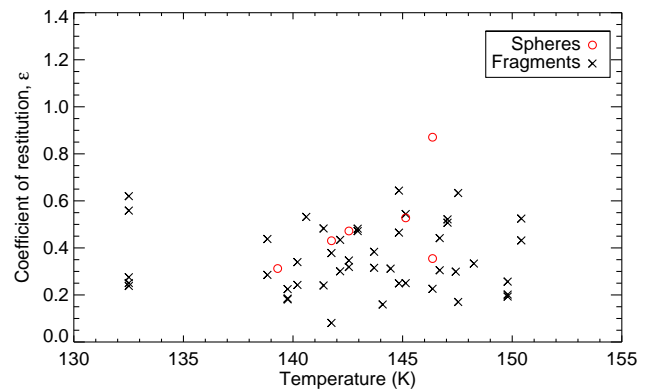}
      \caption{Coefficient of restitution as a function of temperature. There is no correlation between the two parameters.
      }
         \label{FigCOR_T}
   \end{figure}

As it is not possible to cool the experiment with liquid nitrogen while it is onboard the aircraft, the temperature inside the chamber gradually increases as the flight progresses. Fig.~\ref{FigTimeTemp} shows this increase for one flight, in this case the temperature ranged from 131 to 160 K. The temperature increases at a rate of 22.5 K h$^{-1}$ for the first 20 minutes and at a rate of 7.0 K h$^{-1}$ for the remainder of the flight.
Fig.~\ref{FigCOR_T} shows the coefficient of restitution as a function of temperature inside the chamber. There is no correlation between the two parameters (correlation coefficient of 0.02 for the fragments and 0.57 for the spheres); temperature does not affect the coefficient of restitution over this temperature range. This is unsurprising as this temperature range is too low for surface melting effects to come into play which would be expected to reduce the coefficient of restitution.

This graph also indicates that no significant particle frosting takes place either over the course of the experiment or during loading. Since temperature is linearly related to time during a flight, such plots are also a reasonable guide to time dependent effects (if they exist) in the data. If frost accretion during the experiment were a key factor affecting our results then collisions happening later in a flight would occur between particles with the potential to be exposed to more water vapour so carrying a thicker frost layer, and higher temperatures, with the potential to undergo surface melting (or sublimation) leading to increased surface roughness. Both processes have been previously shown by \citet{Hatzes88} to lower the coefficient of restitution of ice spheres colliding with a target by up to 40 \%. Frosting occurring during particle loading would also manifest as a temperature effect as the geometry of the experiment means that the particles that were loaded into the colosseum first were the last to be collided; hence those particles would have the greatest opportunity for surface frosting. However, temperature (and hence time) dependent effects are not observed in our data; in fact even at a single temperature (time) point, the coefficient of restitution values measured show as wide a range as those across the whole temperature (time) dataset. Therefore, we conclude that significant frosting of the particles does not occur; either during the experiment or during the loading procedure.


\section{Conclusions and future work}
\label{Conclusions and future work}

The main conclusions of our work are as follows:

\begin{enumerate}

\item The experimental setup was capable of colliding millimetre-sized ice particles with a broader range of impact parameters, in contrast to the experiments with centimetre-sized ice spheres which only yielded near central collisions.

\item Sticking does not occur as a collisional outcome. Almost universal bouncing, with small amounts of fragmentation in some cases, was observed.

\item The results presented in this paper show that coefficients of restitution of millimetre-sized ice particles are evenly spread from 0.08 to 0.65 with an average of 0.36. Therefore it does not make sense to use a single value for the coefficient of restitution in models where rough, unfrosted surfaces are predicted; instead using a range of values would give a more accurate picture. This corroborates findings for collisions of 1.5 cm (diameter) spheres \citep{Heisselmann10}. The range of coefficients of restitution found is attributed to the surface roughness of our particles.

\item There is no correlation between coefficient of restitution and impact velocity or impact parameter. The translational energy lost in the collision is largely unaffected by the collisional velocity or whether the collision was head on or glancing.

\item The coefficients of restitution are slightly lower on average than those reported for 1.5 cm (diameter) sized spheres at impact velocities of 0.06-0.22~\ms \citep{Heisselmann10}, which were spread from 0.0 to 0.84, suggesting that there is a velocity effect at play with lower velocities giving a greater limit of coefficient of restitution. It is also possible that this effect could be due to the smaller size of our particles (between 4.7 and 10.8 mm).

\item In all cases, some energy is converted into rotation after the collision. This ranges from 0.08\% to 17\%, whereas the energy which is converted into translation ranges from 4\% to 41\%.

\item Temperature did not affect the coefficients of restitution over the range measured (131 to 160 K).

\end{enumerate}

In future, we plan to repeat these experiments with ice particles composed of amorphous solid water, as this is the type of ice that is likely found in space, in the hope of shedding further light on the puzzle of planet formation. For this to be realised, a technique to produce and handle macroscopic quantities of amorphous ice needs to be developed. As a step towards this, we are currently setting up a cryogenic drop tower in the Braunschweig laboratory, which will allow us to study collisions among ice particles in the size range from $\sim$ 1 mm to $\sim$ 100 mm, under controlled temperature conditions, and for all possible impact parameters in the velocity range between $\sim$ 0.01 \ms and $\sim$ 3 \ms.

\begin{acknowledgements}
We thank the Deutsches Zentrum f\"ur Luft- und Raumfahrt (DLR) for providing us with the parabolic flights and for financial aid through grant nos. 50WM0936 and 50WM1236. We thank the European Space Agency (ESA) for providing us with the ICAPS high-speed camera system during the parabolic flight campaign. D. Hei{\ss}elmann was supported by the Deutsche Forschungsgemeinschaft (DFG) through grant no. BL 298/11-1. We would like to thank the European Community's Seventh Framework Programme LASSIE FP7/2007-2013 (Laboratory Astrochemical Surface Science in Europe) for funding H. J. Fraser and C. R. Hill's participation in the data analysis and interpretation from this work, under grant agreement no. 238258. C. R. Hill thanks the Open University for a PhD studentship; H. J. Fraser thanks SUPA (Scottish Universities Physics Alliance) and The University of Strathclyde for supporting the original experimental work. Last, but not least, we thank all parabonauts (Eike Beitz, Stephan Olliges, Ole Sch{\"u}tt and Bob Dawson) who participated in the parabolic-flight campaign for their help in executing the experiments.

\end{acknowledgements}

\bibliographystyle{aa}

\bibliography{ice}

\end{document}